\def\aap{{A\&A}}

\def\aj{{AJ}}
\def\annrev{{ARA\&A}}
\def\apj{{ApJ}}

\def\mnras{{MNRAS}}
\def\nat{{Nature}}
\def\pasj{{PASJ}}

\def\ptl{{Phil. Trans. London}}
\def\rma{{RMA}}

\def\adhoc{{\it ad hoc}}

\def\spose#1{\hbox to 0pt{#1\hss}} % from Scott Tremaine
\def\ltsim{\mathrel{\spose{\lower.5ex\hbox{$\mathchar"218$}}
     \raise.4ex\hbox{$\mathchar"13C$}}}
\def\gtsim{\mathrel{\spose{\lower.5ex \hbox{$\mathchar"218$}}
     \raise.4ex\hbox{$\mathchar"13E$}}}

\documentclass[cup5b]{caps}
\usepackage{psfig}
\usepackage{graphicx}
\usepackage{amssymb}
\usepackage{ociwsymp1}
\HeadText{J. A. Sellwood and Juntai Shen}

\begin{document}

\pagenumbering{arabic}

\author[]{J. A. SELLWOOD and JUNTAI SHEN\\Rutgers University}

\chapter{The AGN-Disk Dynamics \\ Connection}

\begin{abstract}
Any connection between central activity and the large-scale dynamics
of disk galaxies requires an efficient mechanism to remove angular
momentum from the orbiting material.  The only viable means of
achieving inflow from kiloparsec scales is through gravitational
stresses created by bars and/or mergers.  The inflow of gas in bars
today appears to stall at a radius of few hundred parsec, however,
forming a nuclear ring.  Here we suggest that bars in the early Universe may 
have avoided this problem, and propose that the progenitors of central 
supermassive black holes (SMBHs) are created by gas that is driven deep
into the centers of galaxies by bars in the early stages of disk
formation.  The coincidence of the QSO epoch with galaxy formation,
the short lifetimes of QSOs, and the existence of SMBHs in the centers
of most bright galaxies are all naturally accounted for by disk
dynamics in this model.  The progenitor SMBHs are the seeds for
brighter QSO flares during galaxy mergers.  We present a new study of
bar weakening by central mass concentrations, which shows that bars
are less easily destroyed than previously thought.  An extremely
massive and compact central mass can, however, dissolve the bar, which
creates a pseudo-bulge component in the center of the disk.
\end{abstract}

\section{Introduction}
It now seems that the masses of supermassive black holes (SMBHs) in the
centers of galaxies are strongly correlated with the larger scale
dynamical properties of their host galaxies (Kormendy \& Richstone
1995; Gebhardt et al.  2000; Ferrarese \& Merritt 2000).  Yet an
utterly insignificant fraction of the mass of a galaxy had low enough
angular momentum to create, or accrete directly onto, a SMBH in its
center.  Why the mass of the SMBH should be so closely related to the
properties of its host galaxy is still an open question.

Material orbiting at a galacto-centric radius of a few kiloparsecs,
where most of the baryonic galaxy mass resides, must have its angular
momentum reduced by several orders of magnitude before it becomes of
any relevance to nuclear phenomena.  Thus, any connection between
galaxy dynamics and nuclear activity requires a mechanism to remove
enough angular momentum from the gas to enable it to accrete onto the
SMBH.  Viscous processes are too slow for gas to sink from orbits at
large radii to small, even when augmented by magnetohydrodynamic instabilities
(Sellwood \& Balbus 1999), and significant radial migration requires
gravitational torques.  Spiral waves are weak, and generally do more
churning of the gas than radial transportation (Sellwood \& Binney
2002).  Attention has therefore focused on the gravitational influence
of the strongest non-axisymmetric features: bars in isolated systems
and tides during mergers.

Since bars in galaxies today can reduce the angular momentum of gas in
the disk by little more than one order of magnitude, other processes
would be needed to drive gas originating in the main disk of the
galaxy close enough to the nucleus to accrete onto it (see, e.g.,  Wada 2003).
However, the removal of angular momentum by bars could have been somewhat 
more efficient as galaxies first formed, and we propose a possible 
connection between disk dynamics and early QSO activity.

It has often been noted that bright galaxies were assembled at
about the same time that QSOs flare (e.g.,  Rees 1997), suggesting a
causal connection.  In fact, the luminosity function of X-ray selected
AGNs seems to track the star formation history of the Universe
remarkably closely (Franceschini et al.  1999).  Since QSOs are
believed to reside in the centers of galaxies (e.g.,  Bahcall et al. 
1997; McLure et al.  1999), it is natural to suppose that they formed
there.  Many bright galaxies in the local Universe appear to host
quiescent SMBHs which are assumed to be the fuel-starved
engines of earlier QSO activity (Yu \& Tremaine 2002; Ferrarese 2003).

Thus, a convincing model for the formation and evolution central SMBHs
should offer answers to at least the following questions:
\begin{itemize}
\item Why should QSOs flare during an early stage of galaxy formation?
\item Why are the centers of galaxies the preferred sites for QSOs?
\item What interrupts the fuel supply to limit QSO lifetimes?
\item Why is the mass of the central SMBH related to properties of the host bulge?
\end{itemize}

Here we outline a model that offers dynamical answers to the first
three of these questions, but does not yet answer the fourth.  The
main ideas are: (1) most large galaxies developed a bar at an early
stage of their formation, (2) the central engine is created from gas
driven to the center by the bar, and (3) changes to the galaxy
potential, caused by mass inflow itself, shut off the fuel supply to
the central engine when the mass concentration reaches a small
fraction of the galaxy mass.  Furthermore, the central mass weakens
the bar; we show that complete destruction of the bar creates a
(pseudo-)bulge in the stellar distribution but, as yet, we are unable
to demonstrate that this is a necessary consequence of SMBH formation.  This 
picture was proposed by Sellwood \& Moore (1999); Kormendy, Bender, \& Bower
(2002) argue for a similar idea with a somewhat different emphasis.

The early sections of this paper review the various ingredients that
go into this picture, while the later sections put it together.

\section{Gas Flow in a Simple Bar}
Prendergast (1962) was among the first to realize that a rotating bar
in a galaxy would drive large-scale shocks in the interstellar medium,
which he identified with the straight, offset dust lanes commonly seen
on the leading side of the principal axis of the stellar bar.  His
insight has been amply confirmed in a host of gas-dynamical
calculations reported over the past 40 years.

The gas flow pattern is asymmetric about the axis of the bar, leading
to a net torque between the bar and gas.  The gas loses angular
momentum (to the bar) and energy (in the shocks), which drives it
inward.  Unfortunately, the inflow rate is not easily predicted from
theory or simulations because it depends not only on the mass model
for the galaxy and bar, pattern speed, etc., but is particularly
sensitive to the effective viscosity (i.e.,  numerical method and
parameters and perhaps also the assumed equation of state).  The
reason is that the shock is offset farther from the bar major axis as
the effective viscosity increases, leading to an increased rate at
which the gas loses angular momentum.

It is generally desirable to neglect self-gravity in calculations of
the gas flow pattern in a non-axisymmetric potential arising from the
more massive stellar component.  Self-gravitating, dissipative gas
tends to form massive clumps that are, in reality, disrupted by
energetic ``feedback'' from star formation.  The wide range of spatial
scales makes it impossible for a global simulation of the gas flow to
include the small-scale gas dynamics of star formation and feedback in
any meaningful way --- processes that anyway are not fully understood.
Thus, the simulator must include a number of \adhoc\ rules to add
energy back to the gas, in addition to calculating its self-gravity,
thus making the calculation enormously more expensive in computer time
for a questionable improvement in realism.

Self-gravity in the gas should not, of course, be neglected when the
gas component is more massive, or when the stellar component is not
far from axially symmetric, so that the self-gravity of the gas {\it
creates\/} the non-axisymmetric structure (e.g.,  Wada 2003).
But this is not the regime of bar flow.

The standard work is by Athanassoula (1992), who shows that the gas
builds up in a ring at the inner Lindblad resonance (ILR), if one is
present, but is driven in still closer to the center if there is no
ILR and the bar is strong.  Whether an ILR exists depends on the
degree of central concentration in the galactic mass distribution ---
generally a quite modest bulge component is likely to ensure that an
ILR exists.

The conventional definition of the Lindblad resonance is for nearly
circular orbits in an axisymmetric potential.  The concept can readily
be generalized for barred potentials to the region where the
orientation of periodic orbits switches from parallel to perpendicular
to the bar major axis, which occurs at a radius somewhat interior to
that of the na\"\i ve definition of the ILR (e.g.,  Contopoulos \&
Grosb\o l 1989).  (The perpendicular orbit family may even disappear
in weak bars with little bulge; gas flow without shocks or inflow is
possible in such cases.)  For simplicity, I loosely use the phrase
``ILR ring'' to describe the dense ring that forms in the region where
gas settles onto non-intersecting orbits in the vicinity of the
perpendicular orbit family.

\section{Nuclear Rings}
The general picture of gas inflow down a bar until it stalls at a ring
is supported by observation: a gas-rich nuclear ring is seen in many
barred galaxies, where an enhanced rate of star formation is observed.
Beautiful examples are seen in {\it Hubble Space Telescope (HST)}\ images: 
e.g.,  NGC 4314 (Benedict et al.  1998) or NGC 1512 and NGC 5248 (Maoz et al.  
2001).  See also the paper by Carollo (2003).

Furthermore, mm interferometers are mapping
the CO emission from the nuclear regions of nearby galaxies at
unprecedented resolution (e.g.,  Sakamoto et al.  1999; Regan et al. 
2001; Schinnerer et al.  2002; Sofue et al.  2003).  The survey by Regan
et al.  includes some nice examples of gas accumulating in the centers
of barred galaxies and even a partial ring with a central hole can be
seen in NGC~4258, although the central hole is not detected in M100.
A number of caveats about these data should be noted, however: (1)
variations in excitation and optical depth of the CO lines would mean
that the observed intensity does not perfectly reflect the CO
distribution, (2) estimates of the total gas density depend on the
adopted ratio of H$_2$ to CO, and (3) interferometers frequently detect
only a fraction of the CO flux detected by single dishes; this is
because they are sensitive only to the inhomogeneous component and are
``blind'' to smoothly distributed emission.

The very existence of star-forming gas rings requires that any inflow
interior to the ring drains the ring more slowly than gas arrives from
large radii.  Wada (2003) reviews possible mechanisms that
can drive inflow inside the ILR ring.  We would add that Englmaier \&
Shlosman (2000) suggest that further mild inflow of gas inside the ILR
ring could be achieved through globally driven sound (or pressure)
waves.  This suggestion is not supported by the {\it HST}\ images, however,
which often reveal a multi-arm dust distribution that must be caused
by other mechanisms.  It is likely that spiral features are created by
self-gravity in the star-gas mixture and that the behavior inside the
ILR, where the quadrupole field of the bar is weak, may not be so
different from that in the nuclear regions of unbarred galaxies.

\section{Double Bars}
While not directly relevant to the main theme of this paper, the
recent discovery of double bars deserves a mention.  These are too
striking, and the isophotal twists are too great, to be simply the
manifestation of a triaxial ellipsoidal light distribution viewed in
projection.  Erwin \& Sparke (2002) find them to be ``surprisingly
common''; they are seen in at least 25\% (perhaps 40\%) of early-type
barred galaxies.  The random distribution of angles between the inner
and outer (or main) bars strongly suggests separately rotating
components.

The inner bar probably lies within the ILR of the primary and is some
10\% to 15\% of length of the primary bar.  The origin and dynamics of
double bars is not well understood at present, however (see, e.g., 
Maciejewski \& Sparke 2000; Heller, Shlosman, \& Englmaier 2001).

It is known that the gas flow pattern is {\it not\/} simply a scaled-down 
version of that in the principal bar; there are no offset dust
lanes (Regan \& Mulchaey 1999; Shlosman \& Heller 2002) and inflow may
even be inhibited (e.g.,  Maciejewski et al.  2002).

\section{Do Bars Feed AGNs?}
Most studies (e.g.,  Ho, Filippenko, \& Sargent 1997) find no significant 
excess of AGN activity in galaxies with bars over their unbarred counterparts,
although enhanced star formation in the circumnuclear environment has
long been established (Hawarden et al.  1986).  Erwin \& Sparke (2002)
also find no evidence for excess activity in double barred galaxies.

\begin{figure}
\centering
\psfig{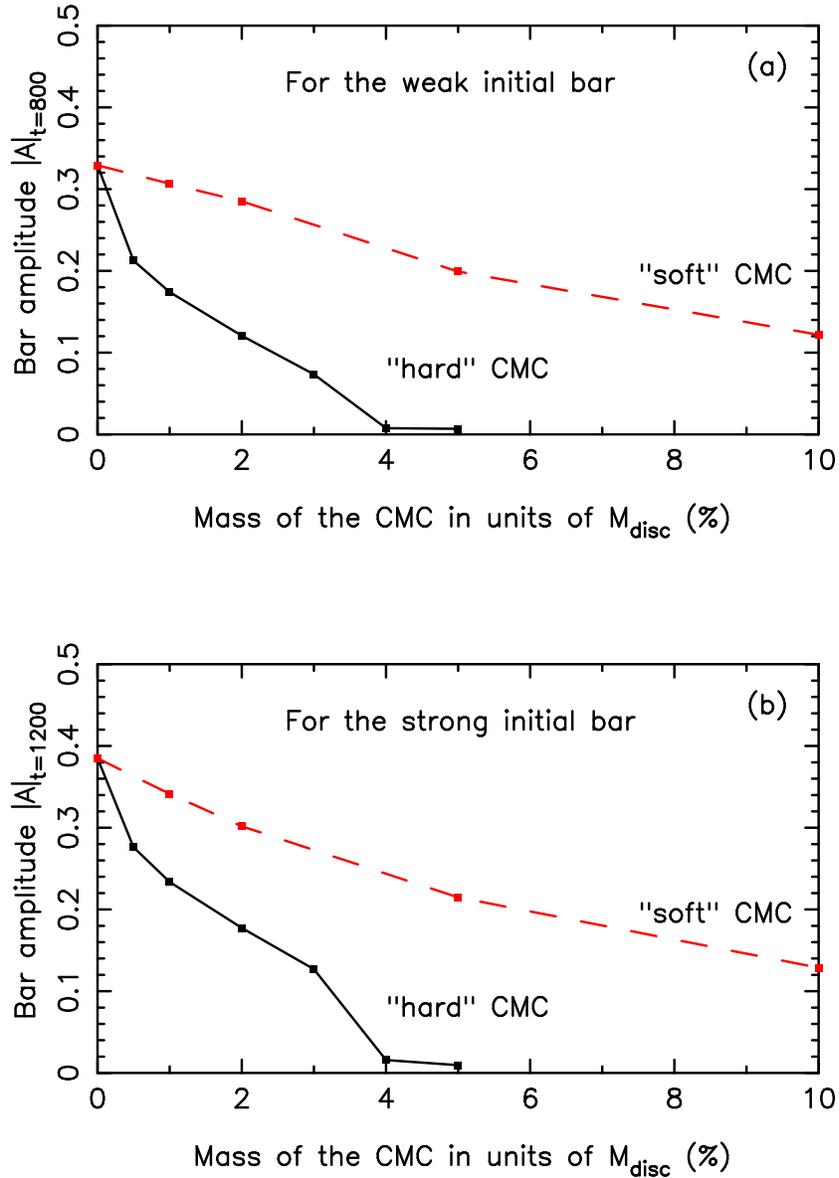}
\caption{The amplitude of the bar at a fixed time ($\sim 6\;$Gyr) after
the introduction of CMCs having a range of masses, for both weak ({\it a}) 
and strong ({\it b}) initial bars.  The bar is completely destroyed only 
by massive, dense CMCs.}
\label{robust}
\end{figure}

However, Laine et al.  (2002) claim weak evidence ($\sim 2.5\sigma$)
for an excess of Seyfert activity in barred galaxies, particularly of
later Hubble types.  We do not find their result compelling, because it
is based on binary binning; there is a continuum both of bar
strengths, and of Seyfert activity levels, and the fractions in each
bin (Seyfert or non-Seyfert, and barred or not) must depend on where the
dividing lines are drawn.  Furthermore, visual classification of
barred versus unbarred is subjective.  It would be better to look for a
correlation between a quantitative estimate of bar strength (e.g., Abraham \& 
Merrifield 2000; Buta \& Block 2001) and some index of ``AGN activity.''

\section{Dissolution of Bars}
Many studies (e.g.,  Pfenniger \& Norman 1990) claim that central mass
concentrations (CMCs) will dissolve bars.  But there have been no
previous systematic studies to determine what would be required to
dissolve bars, either partially or completely; some (e.g.,  Friedli
1994; Hozumi \& Hernquist 1999) have even claimed that very small CMCs
will dissolve bars on a moderate time scale.

\begin{figure}
%\centering
\hskip 0.5truein
\psfig{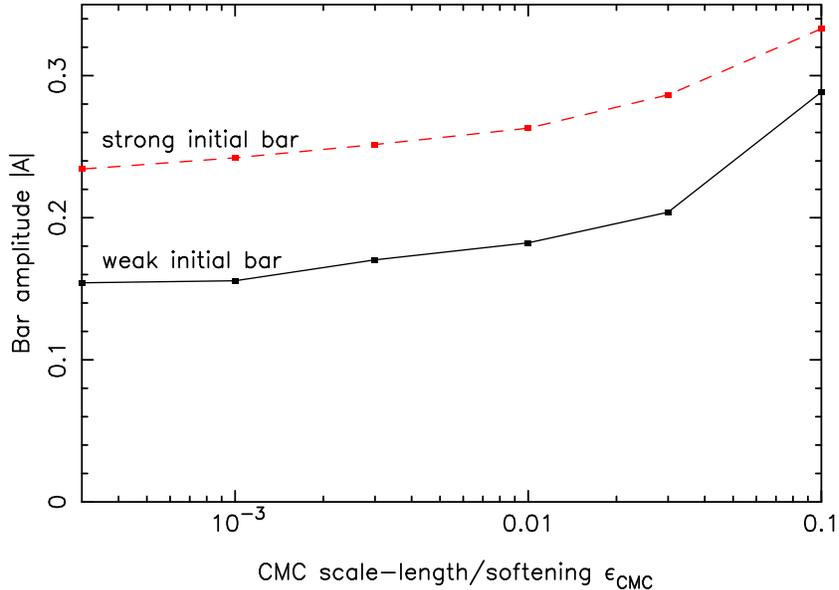}
\caption{The amplitude of the bar at a fixed time ($\sim 3\;$Gyr)
after the introduction of CMCs having a range of central
concentrations, for both strong and weak initial bars.  The softening
length $\epsilon_{\rm CMC}$ is in units of the exponential scale
length of the original disk.  Bars are weakened more by dense CMCs
than by diffuse ones.}
\label{dense}
\end{figure}

Yet bars with CMCs are common, a fact that has led to speculation
that the observed bar fraction may indicate the ``duty cycle'' of bars
in galaxies that repeatedly dissolve and form again (e.g.,  Bournaud \&
Combes 2002).  Since just one cycle of bar formation and destruction
leads to a dynamically very hot disk, such a scenario demands
prodigious infall of fresh gas before a disk could become responsive
enough to form a new bar.  If bars really were fragile, this daunting
requirement might need to be invoked, but, fortunately, we now know
that real bars can survive with realistic CMCs.

\begin{figure}
\centering
\psfig{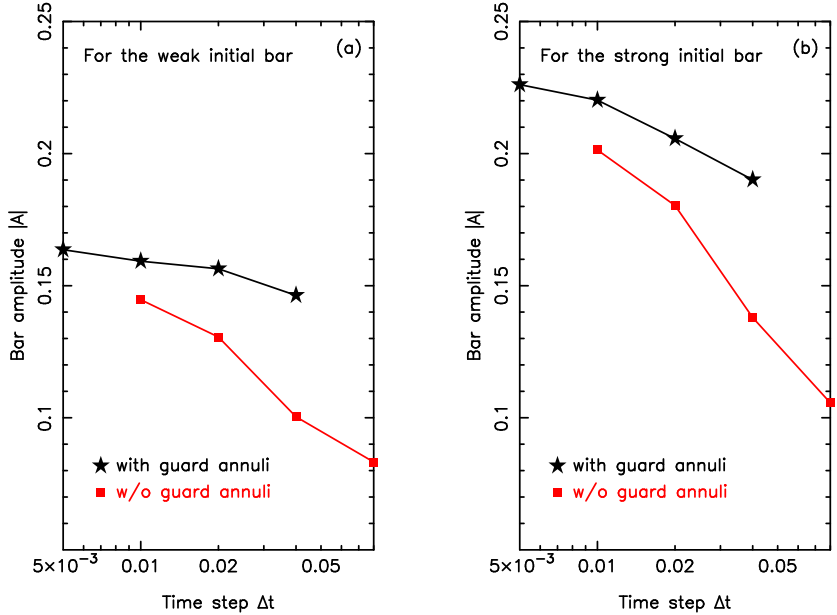}
\caption{The amplitude of the bar at a fixed time ($\sim 3.5\;$Gyr)
after the introduction of a 2\% disk mass CMC as the time step is
varied (lower curves).  The upper curves show the amplitude when the
time step for orbit integration is sub-divided in nested guard annuli
around the CMC.}
\label{timestep}
\end{figure}

We have made the first systematic study of the effect of a CMC on the
survival of the bar and our two major findings are illustrated in
Figures \ref{robust} and \ref{dense}.  We use the amplitude of the
$m=2$ Fourier component of the particle distribution, relative to the
axisymmetric term, as a measure the bar amplitude.  (See Shen \&
Sellwood 2003 for more details.)

Figure \ref{robust} shows that bars are more robust than some previous
studies have suggested.  The bar is totally destroyed only when the
CMC is very dense with a mass $\gtsim 4\%$ of the {\it disk\/} --- a
less massive or more diffuse central mass weakens the bar, but does
not totally destroy it within $\sim6\;$Gyr.  Figure \ref{dense} shows
the trend in bar amplitude at late times as the radial scale of the
Plummer sphere used to model a CMC of fixed mass is varied.  Dense
CMCs are much more destructive than are diffuse CMCs, with a
suggestion that the trend asymptotes to a limit as the size shrinks
toward a point mass.

The critical value of $\sim4\%$ of the {\it disk\/} mass needed for
rapid bar dissolution by a pointlike CMC is enormously larger than
the observed masses of central SMBHs.  Gas accumulation at an ILR ring,
for example, would also have to be quite unreasonably massive ($\gtsim
10\%$ of the disk mass) to threaten the survival of a bar.  Thus,
neither current central SMBHs nor gas concentrations pose a significant
threat to the survival of bars today.

Our results are based on very high quality $N$-body simulations, which
have been extensively checked.  The results shown are for a regime
well clear of significant dependence on the numerical parameters.  We
have found it essential to pay particular attention to the time step.
Figure \ref{timestep} shows that poor orbit integration can cause an
erroneous decay of the bar when too long a time step is used in the
vicinity of a CMC.  We integrate the orbits of particles in the
vicinity of the CMC with time steps that are repeatedly halved (as
many as nine or ten times) in a set of nested guard annuli around the
CMC.  It is likely that previous work overestimated the bar decay
caused by CMCs because of inadequate care in orbit integration.

Not only does the substantial bar fraction in real galaxies suggest
that CMCs pose little threat to bars, but the theoretical picture of
scattering of stars on box orbits by SMBHs (e.g.,  Gerhard \& Binney
1985) really does not apply to bars with rapidly tumbling figures
where most orbits are loop-type ($x_1$) that avoid the center.  A more
likely mechanism for the destruction of bars by massive, dense CMCs is
through the breakdown of regular orbits (e.g.,  Norman, Sellwood, \& Hasan 
1996).  It is perhaps not too surprising that a large, dense mass is required 
to create a sufficiently extensive chaotic region.

It is very hard to imagine that bar destruction by this mechanism
could be achieved more than once in any given galaxy.  The formation
of a new bar is more difficult, because the disk has become both hot
and has acquired a dense center (which inhibits one of the two
possible bar forming mechanisms --- see next section), but is thought
to be possible.  However, the dense center would cause the inflow in
the new bar to stall at the ILR, preventing the growth of a compact
CMC that could threaten its survival.

\section{Formation of Bars --- a Tale of Two Halos}
There are two known mechanisms through which a disk galaxy could
acquire a bar: (1) a global instability or (2) orbit trapping.  The
path adopted depends on the mass distribution.

We find a global instability, with no ILR (initially), when the
density profile has a large, quasi-uniform core.  The global
instability occurs on an orbital time scale, and therefore gives rise
to a bar immediately in any disk that finds itself in an unstable
regime.  Such a situation could arise as the mass of the disk
increases as primordial gas cools in a protogalactic halo and settles
into rotational balance.

Orbit trapping, on the other hand, is the only viable mechanism to
form a bar in a galaxy with a steep, inwardly rising density profile.
This mechanism is also generally quite fast (see Lynden-Bell 1979 for
an alternative) but requires a trigger, such as a mild tidal
interaction (e.g.,  Noguchi 1996) or strong spiral patterns caused by
the build up of significant quantities of new, low-velocity dispersion
material, in the disk (Sellwood 1989; Sellwood \& Moore 1999).  But
isolated disks having dense centers are able to survive for long
periods without making bars (Toomre 1981; Sellwood \& Evans 2001).

\begin{figure}[!p]
\centering
\psfig{figure=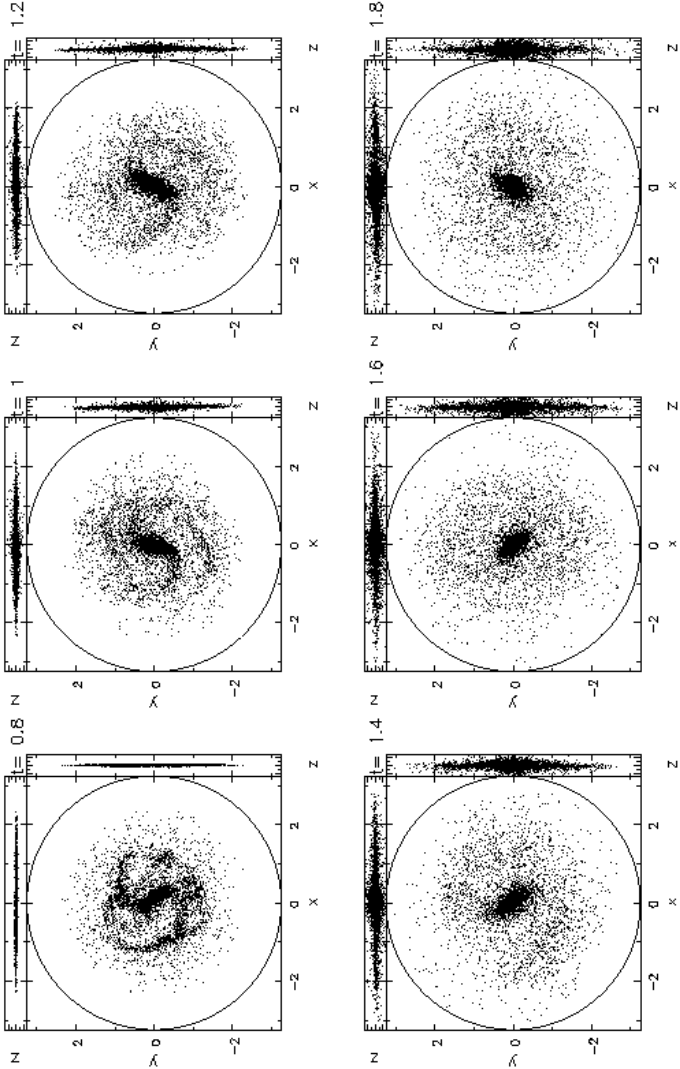,width=\hsize,angle=0,clip=} % run1900
\caption{({\it a}) The evolution of the disk components in two simulations with
different halos.  Times are in Gyr and lengths in kpc.  A
simulation with a soft-core halo profile ($\rho_{\rm halo}
\rightarrow\;$constant as $r\rightarrow0$).}
\label{twohalos}
\end{figure}

As galaxies form, the mass distribution is dominated by the dark matter halo at
first.  If the halo has a large, low-density core, we should expect a
bar to form once the disk mass begins to dominate in the center.  Bars
that may develop in halos with density profiles that rise steeply
toward the center are formed through orbit trapping in the early,
gas-rich disk.  The evolution in the two cases is shown in Figure
\ref{twohalos}, and the extensive differences are summarized in Table
\ref{bardif}.

\addtocounter{figure}{-1}

\begin{figure}[!p]
\centering
\psfig{figure=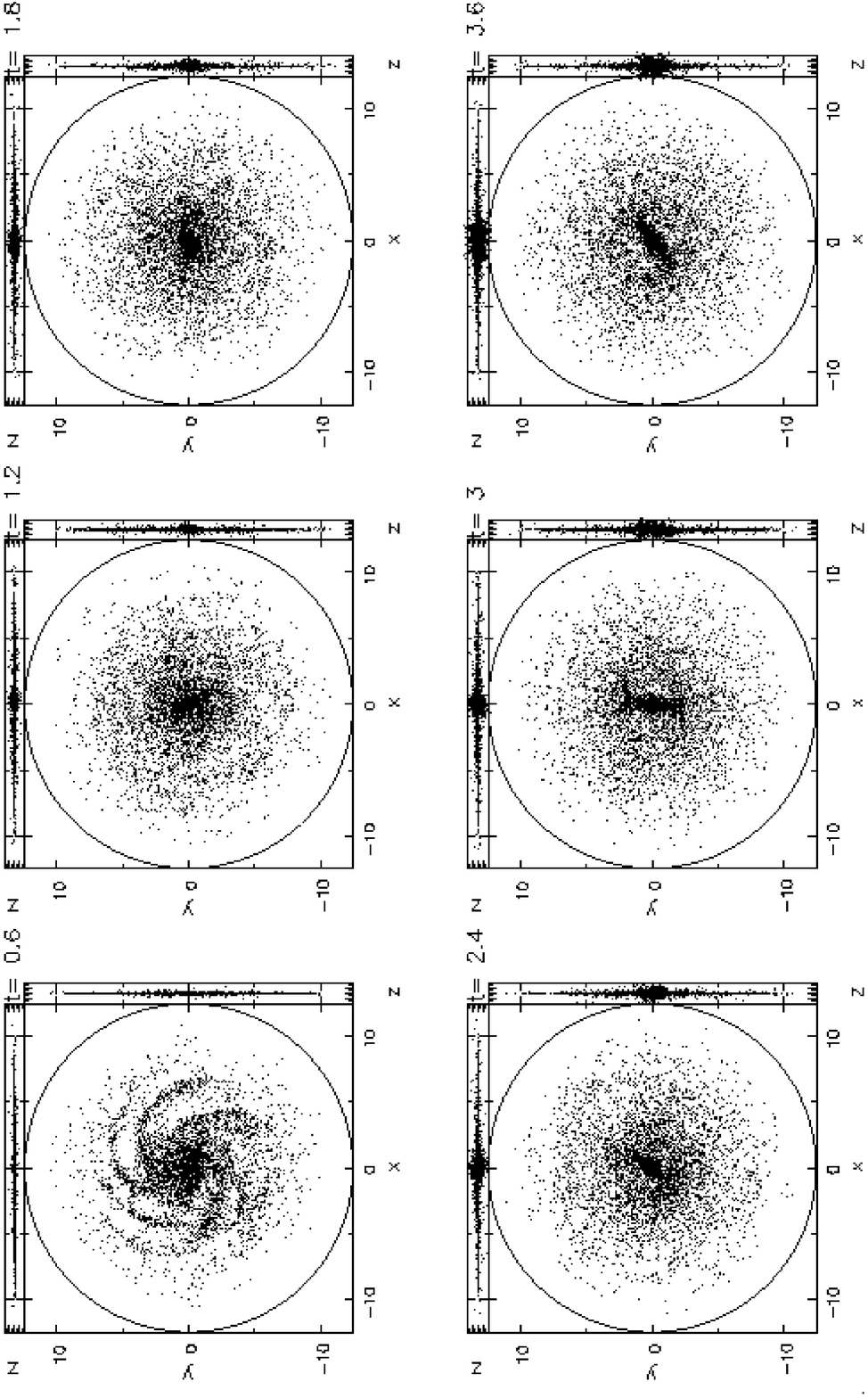,width=\hsize,angle=0,clip=} % run1821
\caption{({\it b}) As in ({\it a}), but for a halo with a cusped density
profile ($\rho_{\rm halo} \propto r^{-1}$ as $r\rightarrow0$).}
\end{figure}

Fig. \ref{twohalos}{\it a}\ shows that a large bar forms in the disk when
the halo has a soft core ($\rho_{\rm halo} \rightarrow\;$constant as
$r\rightarrow0$).  Soon thereafter it undergoes a collective buckling
(aka firehose) instability, caused by the anisotropy of the velocity
distribution between the in-plane and vertical velocity dispersions
(Toomre 1966; Raha et al.  1991; Merritt \& Sellwood 1994).  The
saturation of this instability converts some of the radial motion into
vertical, causing the bar to weaken and to become thicker than the
disk from which it formed (see also Combes \& Sanders 1981).  It seems
likely that any gas in the bar region would be driven still closer to
the center by this event, although we are unaware of any simulations
of gas in a buckling stellar potential.

The bar in the cusped halo ($\rho_{\rm halo} \propto r^{-1}$ for small
$r$), on the other hand, is short at first and grows in length over time 
(Fig. \ref{twohalos}{\it b}).  It also thickens somewhat, but because it
grows gradually, it does not undergo a severe bending convulsion at
any stage.

\clearpage

Gas can be driven deep into the center in the soft-core case, where no
ILR is present (initially), and then probably be further compressed by
the buckling event.  These two successive dynamical instabilities in
the gas-rich early stages ``naturally'' cause a large accumulation of
gas in a small volume close to the center on a short time scale.  A
large concentration of gas is widely believed to be a prerequisite
for the growth of a central SMBH (see Shapiro 2003), but
some other mechanism may be needed to remove more angular momentum
before gas reaches the density required.  By contrast, the initial
ILR in the cusped-halo case will halt gas inflow at some distance from
the center, and the gentle flexing that thickens the bar is unlikely
to have much effect on its radial distribution.

\begin{table}
\begin{center}
\caption{Differences between the Bars Formed in Two Different Halos}
\begin{tabular}{r|cc}
\hline \hline
 &   Soft core  &  Cusped halo \cr
\hline
Initial bar & large & short \cr
ILR & not initially & yes \cr
Major buckling event & yes & no \cr
Later evolution & smaller and weaker & growing \cr
\hline \hline
\end{tabular}
\label{bardif}
\end{center}
\end{table}

The mass of gas accumulated in the center does not have to be very
large ($\ltsim 2\%$ of the disk mass) to change the global potential
in the soft-core halo by enough to introduce an ILR (Sellwood \& Moore
1999).  As soon as the ILR is created, further gas supply to the
center is shut off.

Thus dynamical evolution of the gas in the soft-core halo case
suggests a picture for the origin of QSO activity that has several
appealing features: it creates massive concentrations of gas in galaxy
centers at the time of galaxy assembly, with a mass perhaps related to
the bulge (see below), and a reason the fuel supply is shut off
quickly.  Note also that the SMBH mass need not be as large as that of
the CMC from which it is created, indeed it would be surprising if it
were; a larger fraction will form stars and some may be expelled in a
wind.

Attractive as it is, such a picture is incomplete for two very obvious
reasons: (1) not all SMBHs are in barred galaxies, and (2) it seems to
be established that the brightest QSOs are found in merging or
elliptical (i.e.,  post-merger) galaxies (e.g.,  McLure et al.  1999).

Taking the second point first: It seems likely that at least one
galaxy in a merging pair must already host a SMBH in order to make a
bright outburst.  A bar, and any associated ILR barrier, will be
destroyed in the merger, allowing plenty of fresh fuel to be driven
inward --- this time by the non-axisymmetric forces from the ongoing
merger.  Since the QSO is reignited, and the mass of the SMBH
increases from its previous value, we must expect the brighter QSOs to
be found in merging, or post-merger galaxies.

\section{SMBHs in Non-barred Galaxies}
The other problem is the absence of bars in some galaxies with SMBHs.
Gas inflow in the early bar creates a CMC from which the SMBH is
made.  Since there is no initial ILR to stall the inflow, the gas
concentration will be compact and the bar will be destroyed quickly if
such a CMC exceeds $\sim 4$\% of the disk mass (Figure \ref{robust}).
However, it is likely that an ILR will form well before the central
mass reaches this value, limiting the maximum compact mass that can
be achieved.

The bar is weakened substantially by a CMC of $\sim 2\%$ of the disk
mass (Fig. \ref{robust}), making it more vulnerable to other
destruction mechanisms.  Sellwood \& Moore (1999) found that ongoing
spiral activity in the outer disk, which is not included in our
present simulations, could either complete the destruction of the bar
or cause it to grow again.  Bars can also be destroyed in minor
mergers (e.g.,  Gerin, Combes, \& Athanassoula 1990).

\begin{figure}
\centering
\psfig{figure=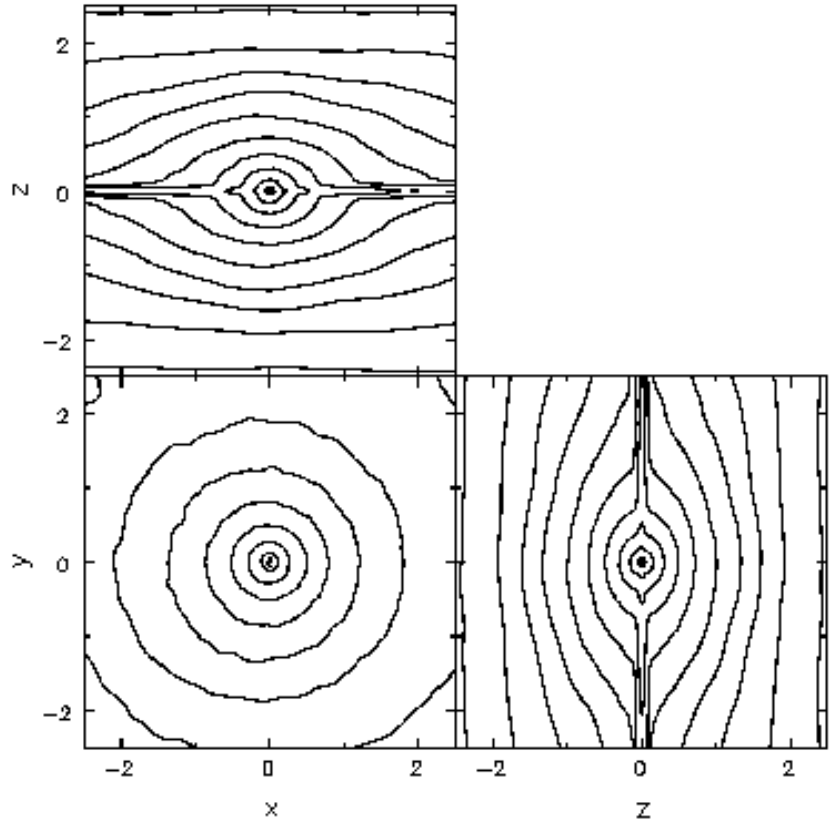,width=1.0\hsize,angle=0,clip=}
\caption{Contours of the projected density of the inner particle
distribution in 3 projections after a strong bar is fully destroyed.
Distances are reckoned in units of the exponential scale length of the
original disk.  Notice the bulgelike feature in the edge-on views.}
\label{bcont}
\end{figure}

\begin{figure}
%\centering
\hspace{-1.5cm}
\psfig{figure=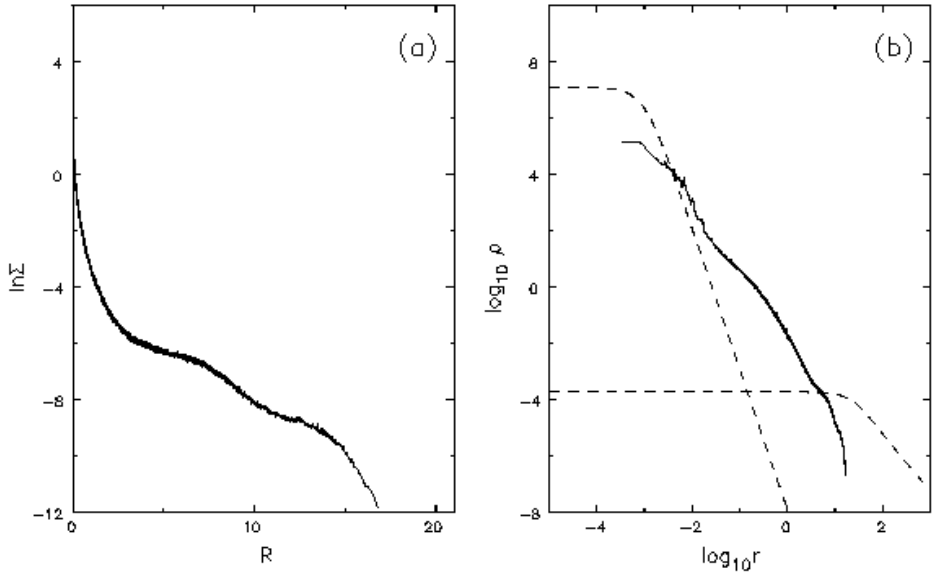,width=1.1\hsize,angle=0,clip=}
\caption{The density profile of the particle distribution (solid
curves) after a strong bar is fully destroyed.  Panel ({\it a})
shows the face-on projected surface density of the disk and bulge,
while panel ({\it b}) shows the volume density.  The dashed
curves in panel ({\it b}) show the density profiles of the central mass
and the halo.  The unit of distance is again the scale length of the
{\it original\/} exponential disk.}
\label{bprof}
\end{figure}

Van den Bergh et al.  (1996) suggest a deficiency of bars in all
galaxies at $z>0.5$, although the claim is disputed (e.g.,  Sheth \&
Regan 2003).  There is no doubt that some bars can be missed in blue
images (Eskridge et al.  2000; Dickinson 2001), but van den Bergh
et al.  (2002) vigorously defend the deficiency.  New data from the ACS
on {\it HST}\ will soon settle the question.  If the deficiency is real, a
second bar must be formed later to account for their present
abundance.

\section{Pseudo-bulges}
An appealing by-product of bar dissolution is that the stars which
were in the bar form an axisymmetric (pseudo-)bulge component in the
galaxy center.  Figures \ref{bcont} and \ref{bprof} are made from the
particle distribution at a time after a strong bar has been totally
destroyed by a 5\% central mass.  The projected density distribution
(Fig. \ref{bcont}), which started exclusively in a disk, reveals a
distinct central bulge that is only slightly flattened in the inner
parts.  The projected density profile of the model at this time
(Fig. \ref{bprof}{\it a}) suggests two separate components, even though
all particles started in a single disk component.  Two distinct
components are also seen in the volume density profile (Fig.
\ref{bprof}{\it b}), which also shows the profiles of the rigid central mass
and the halo components.  (The spherical average used in this plot is
appropriate for the bulge, but obviously not for the disk.)

Norman et al.  (1996) show that such a pseudo-bulge has a high degree of
rotational support, and the velocity field is cylindrically symmetric,
as observed (Kormendy 1993).

\section{Halo Mass Profiles}
It would be inappropriate here to review the current controversy over
the cosmologically predicted dark matter halo density profiles in
galaxies.  The solution to the serious discrepancy between the
predicted and the observed density profiles is far from clear at
present.  But we would like to stress that the predicted cuspy mass
profiles would force an ILR in every bar when it first formed, which
would preclude the formation mechanism for QSOs proposed here.

\section{Relation to Other Models of SMBH Formation}
Following Toomre \& Toomre (1972), many workers (e.g.,  Kauffmann \&
Haehnelt 2000; Di Matteo et al.  2003; Hatziminaoglou et al.  2003)
argue that SMBHs form in mergers, which characterize galaxy formation
in cold dark matter (CDM) Universes (e.g.,  Wechsler et al.  2002).  Such an 
idea has difficulty accounting for SMBHs in galaxies that have long avoided
significant mergers; examples include the Milky Way and those
possessing pseudo-bulges (Carollo 2003).

The picture proposed here operates in every galaxy in which the disk
is massive enough to form a bar through a global instability, and
therefore accounts for the existence of SMBHs in every $L \gtsim L_*$
galaxy.  However, the two proposals are not mutually exclusive, and we
suggest that the seed SMBHs formed by our mechanism are required to be
present in the merging fragments in order to produce a bright
outburst.  The larger SMBHs in giant elliptical galaxies must have grown
substantially during the mergers that formed them.

Furthermore, the idea of angular momentum removal by bar formation
would be of help in making SMBHs by direct collapse, as discussed by
Bromm \& Loeb (2003), for example.

\section{Conclusions}
Inflow in barred galaxies today is arrested at a nuclear ring, which
is identified as the inner Lindblad resonance (ILR) of the bar.
Theoretical predictions of possible further inflow interior to the
ring are not yet mature, but the evident nuclear rings in many {\it HST}\
images and mm interferometer maps imply that gas drains from the ring
at a much slower rate than it arrives from larger radii.  It is likely
the inflow rate inside the ILR ring is little different from that in
the nuclear regions of unbarred galaxies.

If bars can form without ILRs in the early Universe, then gas can be
driven farther into the centers of galaxies.  Gas in a bar-unstable,
young galaxy disk is driven into the center by two successive
instabilities: the usual bar instability, followed by a buckling
instability.  The gas concentration in these gas-rich early stages is
likely to be large, so these instabilities deliver a substantial mass
of gas to within $\ltsim 100\;$pc of the galaxy center.  While bar
flow removes a large fraction of the angular momentum from the gas,
other mechanisms, as yet unclear, are required to reduce it still
further before an SMBH can form.  We simply assume that an SMBH is
created in every galaxy in which the disk becomes massive enough to
become bar unstable.

The gas inflow itself alters the mass distribution enough to create an
ILR where none previously existed, shutting off further gas supply to
the nuclear region.  It is likely that only a fraction of the gas
driven into the center collapses to form the SMBH; probably a larger
fraction will form stars, while some might escape in a wind.  Any
remaining gas that can accrete onto the SMBH will power a
low-luminosity QSO for a while.

This picture of SMBH formation accounts for the coincidence of the QSO
epoch with that of galaxy formation, the fact that SMBHs are found only
in the centers of galaxies, and the short lifetime of QSO activity,
because inflow is shut off by the creation of an ILR.  The modest SMBHs
formed in this way are the seeds for stronger outbursts that must
occur when fresh gas is supplied during mergers.

Sellwood \& Moore (1999) originally hoped that a further consequence
of the initial CMCs that make the SMBHs would be the
destruction of the bar to make a small bulge.  Unfortunately, we now
find that the CMC mass needed to dissolve the bar entirely is larger
than could be assembled by gas inflow down the bar; the ILR probably
intervenes to shut off the inflow at about half the critical mass for
bar destruction.  Bars can still be destroyed later, creating a
bulgelike component, but this event is decoupled from the initial SMBH
formation, which implies that our picture probably cannot hope to
offer a simple understanding of SMBH systematics.

Our model of SMBH formation requires the dark matter halos, in which
the galaxies form, to have large cores --- the instabilities that 
create the central gas concentration do not occur if the disk forms in
a cusped halo.  As there are a number of lines of evidence to suggest
that the cusped halos predicted in CDM cosmology are not present in
real galaxies, we do not regard this requirement to be at all
unrealistic.  It does, however, preclude predictions of the epoch and
rate of QSO activity until whatever is wrong with the CDM prediction
is corrected.

\section*{Acknowledgments}
We would like to thank Laura Ferrarese for a careful read and
thoughtful comments on a draft of this paper.  This work was supported
by NSF grant AST-0098282 and by NASA grant NAG 5-10110.

\end{document}